\documentclass{article}
 \usepackage[dvipdf]{epsfig}
  \usepackage{color}
\begin{document}

\def\a{{\alpha }}
\def\b{{\beta }}
\def\z{{\zeta }}
\def\be{\begin{equation}}
\def\ee#1{\label{#1}\end{equation}}
\def\d{\textsf{d} }
\def\x{\textsf{x} }
 \def\bx{\mathbf{x} }
 \def\bQ{\mathbf{Q} }
 \def\k{\mathbf{k} }
 \def\bp{\mathbf{p} }
 \def\p{\textsf{p} }
\def\no{\nonumber}
\def\lb{\label}
\def\Ki{{\rm Ki}}
\def\J{\textsf{J} }
\def\no{\nonumber}
\def\lb{\label}
\def\h{\textsf{h} }
\def\e{\textsf{e} }
\def\x{\textsf{x} }
\def\n{\textsf{n} }
\def\D{\textsf{D}}
\newcommand{\ben}{\begin{eqnarray}}
\newcommand{\een}{\end{eqnarray}}


 \title{\bf Diffusion of relativistic gas mixtures \\in gravitational fields}

 \author{Gilberto  M. Kremer\footnote{kremer@fisica.ufpr.br}     
  \\
 Departamento de F\'{\i}sica, Universidade Federal do Paran\'a, Curitiba, Brazil }
 \date{}
 \maketitle


\begin{abstract}
 A mixture of relativistic gases of non-disparate rest masses  in a Schwarzschild metric is studied on the basis of a relativistic Boltzmann equation in the presence of gravitational fields. A BGK-type model equation of the collision operator of the Boltzmann equation is used in order to compute the non-equilibrium distribution functions by the Chapman-Enskog method. The main focus of this work is to obtain Fick's law without the thermal-diffusion cross-effect. Fick's law has four contributions, two of them are the usual terms proportional to the gradients of concentration and pressure. The other two are of the same nature as those which appears in Fourier's law in the presence of gravitational fields and are related with an acceleration and gravitational potential gradient, but unlike Fourier's law these two last terms are of non-relativistic order. Furthermore, it is shown that the coefficients of diffusion depend on the gravitational potential and they become larger than those in the absence of it.
\end{abstract}

 \section{Introduction}

 The research on non-equilibrium properties of relativistic gases by using the Boltzmann equation in gravitational fields is a subject few explored in the literature.
  An important contribution was due to Bernstein \cite{B}, who obtained the constitutive equation for the non-equilibrium pressure of a relativistic gas and the corresponding transport coefficient of bulk viscosity in a Friedmann-Robertson-Walker metric.

 Recently a relativistic gas in a gravitational field was analyzed in order to determine the influence of the gravitational potential gradient on Fourier's law \cite{AL,K} and of the influence of the gravitational potential on the transport coefficients \cite{K}.

 According to \cite{K} the heat flux in Fourier's law has three contributions,  the usual temperature gradient and two relativistic terms. One of them -- proposed by Eckart from a thermodynamic theory \cite{Eck} -- is connected with the inertia of energy and represents an isothermal heat flux when matter is accelerated. The other -- suggested by Tolman \cite{To1,To2} -- requires that in absence of an acceleration a state of equilibrium of a relativistic gas in a gravitational field is achieved only if the temperature gradient is counterbalanced by a gravitational potential gradient.

 An open question that follows  refers to the modification of Fick's law in the presence of gravitational fields and the answer is the aim of this present work.

 In this work we are interested only in analyzing Fick's law in the presence of gravitational fields without the determination of the thermal-diffusion cross-effect and of the heat flux with the corresponding diffusion-thermal cross-effect, which will be subject of a forthcoming paper. As in the work \cite{K} we use the Schwarzschild metric and analyze a relativistic gas mixture of constituents which have non-disparate rest masses.

Here we show that Fick's law in the presence of gravitational fields has four contributions. The usual contributions due to the concentration gradient and pressure gradient, apart from the two that appear in Fourier's law and which are proportional to the acceleration and gravitational potential gradient. However, unlike Fourier's law these two last contributions are not of relativistic order. We have also obtained that the diffusion coefficients depend on the gravitational potential, becoming larger than the ones in the absence of it.

The work is structured as follows. In section 2 we introduce the Schwarzschild metric, the system of Boltzmann equations in the presence of a gravitational fields and the two first moments of the distribution functions with their corresponding balance equations. A BGK-type model of the Boltzmann equation is introduced in section 3, which depends on a reference distribution function, determined from the assumption that the balance equations of the full Boltzmann equation and of the BGK-type model lead to the same production terms. The non-equilibrium distribution functions are calculated in section 4 by using the Chapman-Enskog method. In section 5 the constitutive equations for the diffusion fluxes -- which correspond to Fick's law --  are determined from the non-equilibrium  distribution functions and the diffusion coefficients are obtained. In the last section the main conclusions of the work are stated. We close the work with two appendices. In the first one it is shown how to calculate the production terms of the partial balance equations of the energy-momentum tensors, while in the second the components of the Christofell symbols in a  Schwarzschild metric are given.

 \section{Basic  Equations}

We consider a relativistic gas mixture of $r$  constituents in a Riemannian space with metric tensor $g_{\mu\nu}$ where the particles of constituent $a=1,\dots r$ have rest mass $m_a$ and are characterized by the space-time coordinates $(x^\mu)=(ct,\bx)$ and momenta  $(p_a^\mu)=(p_a^0,\bp_a)$. The length of the momentum four-vector is constant so that
$g_{\mu\nu}p_a^\mu p_a^\nu=m_a^2c^2$, which implies that
\ben\lb{1}
p^0_a=\frac{p_{a0}-g_{0i}p^i_a} {g_{00}},\qquad p_{a0}=\sqrt{g_{00}m^2c^2+\left(g_{0i}g_{0j}-g_{00}g_{ij}\right)p^i_ap^j_a}.
\een

As in the work \cite{K} we shall adopt the isotropic Schwarzschild metric
\ben\lb{2}
 ds^2=g_0(r)\left(dx^0\right)^2-g_1(r)\delta_{ij}dx^idx^j,\qquad
g_0(r)=\frac{\left(1-\frac{GM}{2c^2r}\right)^2}{\left(1+\frac{GM}{2c^2r}\right)^2},\qquad
g_1(r)=\left(1+\frac{GM}{2c^2r}\right)^4,
\een
where $G$ denotes the gravitational constant and $M$ the mass of the spherical source.

In terms of the isotropic Schwarzschild metric (\ref{2}),  eqs. (\ref{1}) reduces to
 \ben\lb{3}
 p_a^0=\frac{p_{a0}}{g_0}\qquad p_{a0}=\sqrt{g_0}\sqrt{m_a^2c^2+g_1\vert\bp_a\vert^2},\qquad \sqrt{-g}=\sqrt{g_0g_1^3}.
 \een
where $g=\det(g_{\mu\nu})$.

The components of the four-velocity in the isotropic Schwarzschild metric  are
 \be
 U^\mu=\left( \frac{c}{\sqrt{g_0-v^2/c^2}},\frac {\bf v}{\sqrt{g_0-v^2/c^2}}\right),
 \ee{4}
which in a comoving frame $(\bf v=0)$ reduces to $U^\mu=\left({c}/{\sqrt{g_{0}}},\bf{0}\right)$.

A state of the relativistic mixture of $r$ constituents in the phase space spanned by the space-time and momentum coordinates  is described by the set
of one-particle distribution functions
$f({\bf x},{\bf p}_a,t)\equiv f_a, \, (a=1,2,\dots,r)$
such that $f({\bf x},{\bf p}_a,t) d^3x d^3p_a$
gives at time
$t$, the number of particles of constituent
$a$ in the volume element
$d^3 x$ about  ${\bf x}$ and with momenta in the range
$d^3p_a$ about ${\bf p}_a$.

In the presence of  gravitational fields the one-particle distribution function of constituent $a$ satisfies the Boltzmann
equation (see e.g. \cite{CK})
\be
p_a^\mu \frac{\partial f_a}{\partial x^\mu}-\Gamma_{\mu\nu}^ip_a^\mu p_a^\nu
\frac{\partial f_a}{\partial p_a^i}
=\sum_{b=1}^r\int(f_a'f_b'-f_a f_b)F_{ba}\sigma_{ab}
\,d\Omega \sqrt{-g}{d^3p_b\over p_{b0}}.
\ee{5}
Here  $\Gamma_{\mu\nu}^i$ are the Christoffel symbols, $F_{ba}=\sqrt{(p^\mu_ap_{b\mu})^2-m_a^2m_b^2c^4}$ is the so-called invariant  flux, while $\sigma_{ab}$ and $d\Omega$
denote the invariant differential elastic
cross-section and the element of
solid angle that characterizes a
binary  collision
between the particles of constituent $a$ with those of constituent $b$,
respectively.
The binary collision is characterized by the  momentum four-vectors of the particles of two constituents $p_a^\a$ and $p_b^\a$ before collision  and $p'^\alpha_{a}$ and $p'^\alpha_{b}$ after collision, which obey the energy-momentum conservation law $p_a^\alpha+p_b^\alpha=p'^\alpha_{a}+p'^\alpha_{b}.$  Furthermore, the following abbreviations were introduced in (\ref{5}):
\be
f_a'\equiv f({\bf x},{\bf p}_a',t),\quad
f_b'\equiv f({\bf x},{\bf p}_b',t),\quad
f_a\equiv f({\bf x},{\bf p}_a,t),  \quad
f_b\equiv f({\bf x},{\bf p}_b,t).
\ee{6}

The two first moments of the distribution  function are the partial particle four-flow  $N_a^\mu$ and the
partial energy-momentum  tensor
$T_a^{\mu\nu}$ of constituent $a$, which are defined in terms of the one-particle distribution function by
\be
N_a^\mu=c\int p_a^\mu f_a\sqrt{-g} {d^3 p_a\over p_{a0}},
\qquad T_a^{\mu\nu}=c\int
p_a^\mu p_a^\nu f_a\sqrt{-g}
{d^3 p_a\over p_{a0}}.
\ee{7}
By summing the partial quantities we get the particle four-flow  of the mixture $N^\mu=\sum_{a=1}^r N^\mu_a$
and the energy-momentum  tensor of the mixture $T^{\mu\nu}=\sum_{a=1}^r T_a^{\mu\nu}.$

The balance  equations for the
partial particle four-flow and for the partial energy-momentum tensor of constituent $a$ are obtained by multiplying the Boltzmann equation (\ref{5}) by  $ c\sqrt{-g}
{d^3 p_a/p_{a0}}$  and $cp_a^\nu\sqrt{-g}{d^3 p_a/ p_{a0}}$,  and integrating the resulting equations. Hence, it follows
\be
 {N^\mu_a}_{;\mu}=0,\qquad  {T_a^{\mu\nu}}_{;\nu}=\sum_{b=1}^rc\int (p_a^{\prime\mu}-p_a^\mu)f_a f_b F_{ba}
\sigma_{ab} d\Omega\sqrt{-g}{d^3 p_b\over p_{b0}}\sqrt{-g}{d^3 p_a\over p_{a0}},
\ee{8}
where the semicolon denotes the covariant derivative.

 The balance equations for the particle
four-flow and energy-momentum tensor of the mixture are obtained through the sum of (\ref{8}) over all constituents, namely,
$ {N^\mu}_{;\mu}=0$ and  ${T^{\mu\nu}}_{;\nu}=0.$

In this work we are interested in analyzing the diffusion of a relativistic gas mixture in the presence of  gravitational fields. For that end we shall derive only the constitutive equation for the diffusion flux, which corresponds to Fick's law. The diffusion flux $\J^\mu_a$ of the constituent $a$ in
the mixture is introduced through the decomposition of the the particle four-flow as
\be
N_a^\mu=\n_a U^\mu+\J_a^\mu,\quad\hbox{such that}\quad
\J_a^\mu U_\mu=0,
\ee{9}
where $\n_a$ denotes the particle number  density and $U^\mu$ with $U^\mu U_\mu=c^2$ is the four-velocity. In terms of the projector
\be
\Delta^{\mu\nu}=g^{\mu\nu}-{1\over c^2}U^\mu U^\nu,\qquad\hbox{with}\qquad \Delta^{\mu\nu}U_\nu=0,
\ee{10}
the diffusion flux  can be represented -- thanks to (\ref{7})$_1$ -- as
\be
\J_a^\mu=\Delta_\nu^\mu c\int p_a^\nu f_a{d^3 p_a\over p_{a0}}.
\ee{11}

The sum of   (\ref{9}) over all constituents by taking into account the decomposition of the particle four-flow of the mixture $N^\mu=\n U^\mu$
-- with $\n$ denoting the particle number density  of the mixture -- implies that
\be
\n=\sum_{a=1}^r \n_a,\qquad \sum_{a=1}^{r} \J_a^\mu=0.
\ee{12}
Hence,  due
to the  constraint
(\ref{12})$_2$  there exist only $(r-1)$
partial diffusion fluxes that are linearly
independent for a mixture
of $r$ constituents.

\section{BGK-type kinetic model}

In kinetic theory of gases it is usual to  replace the collision operator of the Boltzmann equation by models which are more easy to handle. These model equations are referred generally as BGK-type kinetic models and the first one was proposed independently in 1954 by Bhatnagar, Gross and Krook \cite{BGK} and by  Welander \cite{WE}.
The extension of BGK-type models to relativistic gases  was due to Marle \cite{Mar1,Mar2} and Anderson and Witting \cite{AW}.

The main objective of  this work is to discuss the effect of  gravitational fields in the diffusion of relativistic gas mixtures and to  derive Fick's law in the presence of gravitational fields without the cross effect of thermal-diffusion. For that end we shall use a truncated Grad's distribution function that depends only on the diffusion flux, which will be the only non-equilibrium quantity in this theory. The consideration of the full Grad's distribution function is more involved and this will be the subject of a forthcoming paper, where the heat flux and cross-efects shall be analyzed.

The model equation we shall use is based on the McCormac model for mixtures of non-relativistic gas mixtures \cite{Mc} extended to a relativistic truncated Grad's distribution function. The McCormac model for non-relativistic gas mixtures of non-reacting and reacting gases by using a truncated Grad's distribution function were analyzed in the works \cite{GSB} and \cite{KPS}, respectively.

The truncated Grad's distribution function which takes into account the diffusion flux as the only non-equilibrium quantity is given by (see \cite{KM})
\be
f_a=f_a^{(0)}\left(1-{\J_a^\mu\over \p_a}\,p_{a\mu}\right),
\ee{13}
where $\p_a=\n_a kT$ is the pressure of constituent $a$ and $f_a^{(0)}$  the Maxwell-J\"uttner distribution function
\be
f_a^{(0)}={n_a\over4\pi kTm_a^2cK_2(\zeta_a)}\exp\left(-\frac{U_\mu p_a^\mu}{kT}\right)={n_a\over4\pi kTm_a^2cK_2(\zeta_a)}\exp\left(-\frac{c\sqrt{m_a^2c^2+g_1\vert\bp_a\vert^2}}{kT}\right).
\ee{14}
Here the second equality is written in a comoving frame, where $U^\mu=\left({c}/{\sqrt{g_{0}}},\bf{0}\right)$.
Above $T$ denotes the temperature of the mixture and it was supposed that all constituents are at the same temperature. Furthermore,  $k$ is the Boltzmann constant and $\zeta_a=m_ac^2/kT$ represents the ratio of the rest energy $m_ac^2$ of a relativistic $a$-particle and the thermal energy of the mixture $kT$. The limiting case of low temperatures $\z_a\gg1$ corresponds to a non-relativistic regime and the limiting case of high temperatures $\z_a\ll1$ to an ultra-relativistic regime. Furthermore, $K_n(\z_a)$ denotes the modified Bessel function of second kind which is given by
\ben\lb{15}
K_n(\zeta_a)=\left(\frac{\zeta_a}{2}\right)^n\frac{\Gamma(1/2)}{\Gamma(n+1/2)}\int_{1}^\infty e^{-\zeta_a y}\left(y^2-1\right)^{n-1/2}\,dy,
\een

The BGK-type kinetic model we shall use reads
\be
p_a^\mu \frac{\partial_\a f_a}{\partial x^\mu}-\Gamma_{\mu\nu}^ip_a^\mu p_a^\nu
\frac{\partial f_a}{\partial p_a^i}=-2\sum_{b=1}^r{m_{ab}\,\nu_{ab}}(f_a-f_{ab}^R),
\ee{16}
where $m_{ab}=m_am_b/(m_a+m_b)$ is the reduced mass, $\nu_{ab}$ is a collision frequency whose inverse is of order of the mean free time and $f_{ab}^R$ is a reference distribution function.
In order to be compatible with the truncated Grad's distribution function the reference distribution function must be written as
\be
f_{ab}^R=f_a^{(0)}\left(1+B_{ab}+C_{ab}^\mu\, p_{a\mu}\right),
\ee{17}
where $B_{ab}$ and $C_{ab}^\mu$ are coefficients to be determined.

In order to determine the coefficients $B_{ab}$ and $C_{ab}^\mu$ we multiply the BGK-type kinetic model (\ref{16}) by
$ c\sqrt{-g}{d^3 p_a/p_{a0}}$  and $cp_a^\nu\sqrt{-g}{d^3 p_a/ p_{a0}}$ and integrate the resulting equations. Hence, it follows
\ben\lb{18a}
{N^\mu_a}_{;\mu}&=&-2\sum_{b=1}^r{m_{ab}\,\nu_{ab}}\,c\int (f_a-f_{ab}^R)\sqrt{-g}{d^3 p_a\over p_{a0}},\\\lb{18b}
{T_a^{\mu\nu}}_{;\nu}&=&-2\sum_{b=1}^r{m_{ab}\,\nu_{ab}}\,c\int p_a^\mu(f_a-f_{ab}^R)\sqrt{-g}{d^3 p_a\over p_{a0}}.
\een
Comparison of the above equations with (\ref{8}) leads to
\ben\lb{19a}
-2\sum_{b=1}^r{m_{ab}\,\nu_{ab}}\,c\int (f_a-f_{ab}^R)\sqrt{-g}{d^3 p_a\over p_{a0}}=0,\\\lb{19b}
-2\sum_{b=1}^r{m_{ab}\,\nu_{ab}}\,c\int p_a^\mu(f_a-f_{ab}^R)\sqrt{-g}{d^3 p_a\over p_{a0}}=\sum_{b=1}^rc\int(p_a^{\prime\mu}-p_a^\mu)f_a f_b F_{ba}
\sigma_{ab} d\Omega\sqrt{-g}{d^3 p_b\over p_{b0}}\sqrt{-g}{d^3 p_a\over p_{a0}},
\een
and from the above equations we can determine the coefficients $B_{ab}$ and $C_{ab}^\a$. Let us  denote the integrals in (\ref{19a}) and (\ref{19b}) as
\ben\lb{20a}
^1\mathcal{I}_{ab}&=&c\int (f_a-f_{ab}^R)\sqrt{-g}{d^3 p_a\over p_{a0}},\qquad
^2\mathcal{I}_{ab}^\mu=c\int p_a^\mu(f_a-f_{ab}^R)\sqrt{-g}{d^3 p_a\over p_{a0}},\\\lb{20b}
^3\mathcal{I}_{ab}^\mu&=&c\int(p_a^{\prime\mu}-p_a^\mu)f_a f_b F_{ba}
\sigma_{ab} d\Omega\sqrt{-g}{d^3 p_b\over p_{b0}}\sqrt{-g}{d^3 p_a\over p_{a0}}.
\een

The integrations of (\ref{20a}) lead to \footnote{For more details on the calculations of  these integrals one is referred to the work \cite{K}.}
\ben\lb{21a}
^1\mathcal{I}_{ab}&=&-{\p_a\over m_a}\left({B_{ab}\over kT}{K_1(\z_a)\over K_2(\z_a)}+{m_a\over kT}C_{ab}^\mu U_\mu\right),\\\lb{21b}
^2\mathcal{I}_{ab}^\mu&=&-\p_a\left[{B_{ab}\over kT}U^\mu+{K_3(\z_a)\over K_2(\z_a)}{m_a\over kT}C_{ab}^\nu U_\nu U^\mu-C_{ab}^\mu-{\J_a^\mu\over\p_a}\right],
\een
thanks to (\ref{13}) and (\ref{17}).
The integral (\ref{20b}) is calculated in the Appendix A for mixtures of non-disparate masses $(m_b\approx m_a)$ and constant differential cross section $\sigma_{ab}$. Its expression is given by
\be
 ^3\mathcal{I}_{ab}^\mu=-\Upsilon_{ab}\left(\frac{\J_{a}^\mu}{\p_a}-\frac{\J_{b}^\mu}{\p_b}\right),
 \ee{22}
 where the following abbreviation was introduced
 \be
 \Upsilon_{ab}^\b=\frac{16\pi \p_a \p_b \sigma_{ab}}{3c   K_2(\z_a)K_2(\z_b)}\left[\frac{7}{\sqrt{\z_a\z_b}}K_3\left(2\sqrt{\z_a\z_b}\right)+\left(2+\frac{1}{\z_a\z_b}\right)K_2\left(2\sqrt{\z_a\z_b}\right)\right].
 \ee{23}

From (\ref{20a}) -- (\ref{22}) we conclude that $C_{ab}^\mu U_\mu=0$, $B_{ab}=0$ and
\be
C_{ab}^\mu={\Upsilon_{ab}\over 2\,m_{ab}\,\nu_{ab}\p_a}\left({\J_a^\mu\over \p_a}-{\J_b^\mu\over \p_b}\right)-{\J_a^\mu\over \p_a}.
\ee{24}
Hence, the reference distribution function (\ref{17}) becomes
\be
f_{ab}^R=f_a^{(0)}\left\{1+\left[{\Upsilon_{ab}\over 2\,m_{ab}\,\nu_{ab}\p_a}\left({\J_a^\mu\over \p_a}-{\J_b^\mu\over \p_b}\right)-{\J_a^\mu\over \p_a}\right]p_{a\mu}\right\}.
\ee{25}

\section{Non-equilibrium distribution function}

For the determination of the non-equilibrium distribution function we shall rely on the Chapman-Enskog method. First we write the one-particle distribution function of constituent $a$ as
$f_a=f_a^{(0)}\left(1+\varphi_a\right)$ where the deviation of the Maxwell-J\"uttner distribution function is considered a small quantity $(\vert\varphi_a\vert<1)$. According to the Chapman-Enskog method the equilibrium distribution function $f_a^{(0)}$ is inserted on the left-hand side of the BGK-type kinetic model (\ref{16}) and the representation $f_a=f_a^{(0)}\left(1+\varphi_a\right)$ on its right-hand side. Hence, by performing the derivatives we get that the deviation of the Maxwell-J\"uttner distribution function becomes
\ben\no
\varphi_a=-\frac{1}{2\sum_{b=1}^r{m_{ab}\,\nu_{ab}}}\bigg\{\frac{p_a^\nu}{n_a}\frac{\partial n_a}{\partial x^\nu}
+\frac{p_a^\nu}{T}\left[1-\frac{K_3(\z_a)\z_a}{K_2(\z_a)}+\frac{ p_a^\tau U_\tau}{kT}\right]\frac{\partial T}{\partial x^\nu}-\frac{p_a^\tau p_a^\nu}{kT}\frac{\partial U_\tau}{\partial x^\nu}
\\\no
-\frac{c^2}{2kT}\frac{d g_1}{dr}\frac{p_a^ip_a^jp_a^k}{U_\tau p_a^\tau}\delta_{ij}\delta_{kl}\frac{x^l}{r}
+\frac{c^2}{k T}g_1\delta_{ij}\Gamma_{\sigma\nu}^i \frac{p_a^j p_a^\sigma p^\nu }{U_\tau p_a^\tau}
+\frac{c}{2\sqrt{g_0}kT}p_a^ip_a^0\frac{dg_0}{dr}\delta_{ij}\frac{x^j}{r}
\\\lb{27}
-\sum_{b=1}^r{\Upsilon_{ab}\over \p_a}\left({\J_a^\mu\over \p_a}-{\J_b^\mu\over \p_b}\right)p_{a\mu}\bigg\}
-{\J_a^\mu\over \p_a}p_{a\mu}.
\een
Note that the deviation of the Maxwell-J\"uttner distribution function $\varphi_a$ is a function of the diffusion flux $\J_a^\mu$, gradients of $n_a, U^\mu, T$ and derivatives of the components of the metric tensor $g_0, g_1$.

\section{Fick's law}

Insertion of the representation $f_a=f_a^{(0)}\left(1+\varphi_a\right)$ together with (\ref{14}) and (\ref{27}) into the definition of the diffusion flux (\ref{11}) and integrating the resulting equation it follows a relationship which can be used to determine Fick's law, namely,
\ben\lb{28}
\sum_{b=1}^r\Upsilon_{ab}\left(\frac{\J_a^\mu}{\p_a}-\frac{\J_b^\mu}{\p_b}\right)=\Delta^{\mu\nu}\frac{\partial\p_a}{\partial x^\nu}-\frac{\n_a\h_a}{c^2}\left[\Delta^{\mu\nu}U^\tau U_{\nu;\tau}
-\frac{1}{1-\Phi^2/4c^4}\Delta^{\mu j}\frac{\partial\Phi}{\partial x^j}\right].
\een
Here $\h_a=m_ac^2 K_3(\z_a)/K_2(\z_a)$ denotes the enthalpy per particle of constituent $a$ and $\Phi=-GM/r$ the gravitational potential.

If we sum (\ref{28}) over all constituents we get the momentum density balance equation for the mixture (see \cite{K} eq.(22))
\ben\lb{29}
\Delta^{\mu\nu}\frac{\partial\p}{\partial x^\nu}-\frac{\n\h}{c^2}\left[\Delta^{\mu\nu}U^\tau U_{\nu;\tau}
-\frac{1}{1-\Phi^2/4c^4}\Delta^{\mu j}\frac{\partial\Phi}{\partial x^j}\right]=0,
\een
where $\p=\sum_{a=1}^r\p_a$ is the pressure and $\h=\sum_{a=1}^r\n_a\h_a/\n$  the enthalpy per particle of the mixture.

Since we have $r-1$ diffusion fluxes which are linearly independents we rewrite (\ref{28})  as a system of $r-1$ equations for the determination of the diffusion fluxes:
\ben\lb{30}
\sum_{b=1}^{r-1}D^{-1}_{ab} J_b^\mu=\Delta^{\mu\nu}\left[\frac{\partial\x_a}{\partial x^\nu}+(\x_a-1)\frac{\partial\ln p}{\partial x^\nu}\right]-\frac{\n_a\h_a-\n\h}{ c^2\p}\left[\Delta^{\mu\nu}U^\tau U_{\nu;\tau}
-\frac{1}{1-\Phi^2/4c^4}\Delta^{\mu j}\frac{\partial\Phi}{\partial x^j}\right].
\een
Here we have used the constraint (\ref{29})
and  introduced the concentration $\x_a=\p_a/\p=\n_a/\n$ of constituent $a$ and the coefficient
\ben\lb{31}
D^{-1}_{ab}=\frac{1}{\p}\left[\sum_{c=1}^r\Upsilon_{ac}\left(\frac{\delta_{ab}}{\p_a}-\frac{\delta_{ar}}{\p_r}\right)-\frac{\Upsilon_{ab}}{\p_b}+\frac{\Upsilon_{ar}}{\p_r}\right].
\een

In  a comoving frame where
\be
\Delta^{00}=0,\qquad \Delta^{ij}=g^{ij}=-\frac{1}{g_1}\delta^{ij}=-\frac{1}{\left(1-\frac{\vert\Phi\vert}{2c^2}\right)^4}\delta^{ij},
\ee{32}
we obtain from (\ref{30}) Fick's law:
\ben\lb{33}
J_a^i=-\sum_{b=1}^{r-1} \n\mathcal{D}_{ab}\delta^{ij}\left\{\left[\frac{\partial\x_b}{\partial x^j}+(\x_b-1)\frac{\partial\ln p}{\partial x^j}\right]-\frac{\n_b\h_b-\n\h}{c^2\p}\left[U^\tau U_{j;\tau}
-\frac{1}{1-\Phi^2/4c^4}\frac{\partial\Phi}{\partial x^j}\right]\right\}.
\een
Here we identify $ \mathcal{D}_{ab}={D}_{ab}/(\n g_1)$ as the diffusion coefficient.

 From Fick's law (\ref{33}) we may infer  that the diffusion flux -- without the consideration of thermal-diffusion effects -- has  four different contributions:
 \begin{enumerate}
   \item a flux due to a concentration gradient   that tends to reduce the non-homogeneity of the mixture;
   \item a flux associated with the pressure gradient  where heavy particles tend to diffuse to places with large pressures as in the case of centrifuges;
   \item a flux related with an acceleration term, since the acceleration acts on particles with different masses;
   \item a flux caused by the gravitational potential gradient. By taking into account that the acceleration term is absent and that the pressure is uniform,  we may infer from (\ref{33}) that the diffusion flux vanishes if the concentration gradient is counterbalanced with the gravitational potential gradient.
 \end{enumerate}

 Let us compare Fick's law with Fourier's law for a single relativistic fluid. According to eq. (33) of \cite{K} we have
 \ben\lb{34}
 q^i=-\lambda \delta^{ij}\left\{\frac{\partial T}{\partial x^j}-\frac{T}{c^2}\left[U^\tau U_{j;\tau}-\frac{1}{1-\Phi^2/4c^4}\frac{\partial\Phi}{\partial x^j}\right]\right\},
 \een
 where $q^i$ denotes the heat flux and $\lambda$ the thermal conductivity coefficient. By comparing (\ref{33}) with (\ref{34}) we infer that the combination of an accelerated term and a gravitational potential gradient has the same structure in both equations. However,  in the case of the heat flux this combination is of relativistic order which is not case of the diffusion flux, since in the non-relativistic limiting case $(\z_a\gg1)$ the expression  $(\n_b\h_b-\n\h)/{c^2\p}$ tends to $(\x_bm_b-\sum_{a=1}^r \x_a m_a)/kT$.
As was pointed out in \cite{K} the acceleration  and the gravitational potential gradient terms, which appear in the heat flux (\ref{34}), were first proposed by Eckart \cite{Eck} and Tolman \cite{To1,To2}, respectively.

In a non-relativistic theory (see \cite{CC} pp. 344 and 345) the diffusion flux is proportional to the so-called generalized diffusion forces whose expression, in our notation, is given by
 \ben\lb{35}
 \mathbf{d}_a=\nabla \x_a+\left(\x_a-\frac{m_a\n_a}{\sum_{b}m_b\n_b}\right)\nabla\ln\p-\frac{m_a\n_a}{\p}\left(\mathbf{F}_a-\frac{\sum_b m_b\n_b\mathbf{F}_b}{\sum_c m_c\n_c}\right).
 \een
 Here, $\mathbf{F}_a$ denotes a force per unit mass that acts on the particles of constituent $a$. Both equations (\ref{33}) and (\ref{35}) depend on the concentration and pressure gradients, but the third term of (\ref{35}) is of different nature. Indeed, it is connected with an acceleration when the forces acting on the particles of different constituents are unequal. However, this term vanishes when one considers  that only gravitational forces are acting on the particles.

 One important point to call attention is that the acceleration term is always connected with the gravitational potential gradient. If we eliminate from (\ref{33}) the acceleration term by the use of the  momentum density balance equation for the mixture (\ref{29}) both contributions disappears and we get
 \ben\lb{36}
 J_a^i=-\sum_{b=1}^{r-1} \n\mathcal{D}_{ab}\delta^{ij}\left[\frac{\partial\x_b}{\partial x^j}+\left(\x_b-\frac{\n_b\h_b}{\n\h}\right)\frac{\partial\ln p}{\partial x^j}\right].
 \een
Note that the terms in brackets reduce to the first two terms of (\ref{35}) in the non-relativistic limiting case. Although this is a common practice in the non-relativistic and special relativistic kinetic theories, here this operation cannot be done, since in this case the dependence on the gravitational potential gradient disappears from the constitutive equation of the diffusion flux and in our opinion this is not physically correct. This fact was also shown in \cite{K}, that the elimination of the acceleration term from Fourier's law leads to a heat flux which is independent on the gravitational potential gradient.

As in the case of a single relativistic gas in the presence of gravitational fields \cite{K} the diffusion coefficient depends on the gravitational potential through the component of the metric tensor $g_1(r)=(1-\Phi/2c^2)^4$. Let us analyze the diffusion coefficient of a binary mixture which follows from (\ref{31}). In this case we have that
\ben\lb{37}
\mathcal{D}_{11}=\frac{\p_1\p_2}{\n\Upsilon_{12}}=\frac{3 c K_2(\z_1)K_2(\z_2)}{16\pi\n\sigma_{12}\left(1-\vert\Phi\vert/2c^2\right)^4}\left[\frac{7}{\sqrt{\z_1\z_2}}K_3\left(2\sqrt{\z_1\z_2}\right)+
\left(2+\frac{1}{\z_1\z_2}\right)K_2\left(2\sqrt{\z_1\z_2}\right)\right]^{-1}.
\een
From the above equation we infer that the diffusion coefficient becomes larger in the presence of gravitational fields. As was pointed out in \cite{K} this increase is very small for stellar objects like Earth and Sun, but it becomes important for more compact stellar objects, like a neutron star.

The diffusion coefficient (\ref{37}) in the non-relativistic limiting case $\z_1,\z_2\gg1$ that corresponds to a binary mixture of constituent with non-disparate masses $(m_1\approx m_2)$ at low temperatures and under the influence of weak gravitational fields reads
\ben\lb{38}
\mathcal{D}_{11}=\frac{3}{32\n\sigma_{12}}\sqrt{\frac{kT}{\pi m_1}}\left(1-\underline{\frac{11}{16\z_1}}+\dots\right)\left(1+\underline{\frac{2\vert\Phi\vert}{c^2}}+\dots\right).
\een
Without the relativistic and gravitational field corrections given by the underlined terms of (\ref{38}) we get the self-diffusion coefficient of a mixture of hard sphere particles, since in this case the differential cross section is given in terms of the relative diameter of the particles at collision by $\sigma_{12}=(\d_1+\d_2)^2/16$ (see e.g. \cite{CC,KB}).

For high temperatures $\z_a\ll1$ and the diffusion coefficient (\ref{37}) in the ultra-relativistic limiting case under the influence of weak gravitational fields becomes
\ben\lb{39}
\mathcal{D}_{11}=\frac{c}{10\pi\n\sigma_{12}}\left(1-{\frac{\z_1^2}{10}}+\dots\right)\left(1+{\frac{2\vert\Phi\vert}{c^2}}+\dots\right).
\een

\section{Conclusions}

The main objective of this work was to show that Fick's law in the presence of gravitational fields has the same structure as Fourier's law for a single constituent, i.e., apart from the well known forces without the presence of the gravitational field, both laws depend on the acceleration and gravitational potential gradient. From the analysis of Fick's law we have concluded: (i) the driving forces in the absence of gravitational fields are the usual terms connected with the concentration gradient and pressure gradient; (ii) the terms related with the acceleration and gravitational potential gradient unlike Fourier's law are of non-relativistic order; (iii) the transport coefficients of diffusion depend on the gravitational potential and they  become larger than the ones in the absence of it.

\appendix

\section{Appendix: Evaluation of $^3\mathcal{I}_{ab}^\mu$}

In order to perform the  integration of
\ben\lb{A1}
^3\mathcal{I}_{ab}^\mu=c\int(p_a^{\prime\mu}-p_a^\mu)f_a f_b F_{ba}
\sigma_{ab} d\Omega\sqrt{-g}{d^3 p_b\over p_{b0}}\sqrt{-g}{d^3 p_a\over p_{a0}},
\een
we introduce the total momentum four-vector $P^\mu$ and the
relative momentum four-vector $Q^\mu$ defined by (see e.g. \cite{St})
 \ben\lb{A2}
 P^\mu\equiv p_a^\mu+p_b^\mu,\qquad P^{\prime\mu}\equiv p^{\prime\mu}_a+p^{\prime\mu}_b,
 \qquad
 Q^\mu=p_a^\mu-p_b^\mu,\qquad
 Q^{\prime\mu}=p^{\prime\mu}_a-p^{\prime\mu}_b.
 \een
  The momentum four-vector conservation law when applied to the above equations lead to
 \ben\lb{A3}
 P^\mu=P^{\prime\mu},\qquad
 P^\mu Q_\mu=(m_a^2-m_b^2)c^2,
 \qquad
 Q^2=P^2-{2(m_a^2+m_b^2)c^2}.
 \een
 Here $P^2=P^\mu P_\mu$ and $Q^2=-Q^\mu Q_\mu$ are the magnitudes
 of the total and relative momentum four-vectors, respectively.
 Furthermore, the inverse transformations of (\ref{A2})   read
 \ben\lb{A3a}
 p_a^\mu=\frac{P^\mu}{2}+\frac{Q^\mu}{2},\qquad
 p_b^\mu=\frac{P^\mu}{2}-\frac{Q^\mu}{2},
 \qquad
 p^{\prime\mu}_a=\frac{P^\mu}{2}+\frac{Q^{\prime\mu}}{2},\qquad
 p^{\prime\mu}_b=\frac{P^\mu}{2}-\frac{Q^{\prime\mu}}{2},
 \een
and the Jacobian of the transformation from $(p^\mu_a,p^\mu_b)$ to $(P^\mu, Q^\mu)$ is 1/8, so that $d^3p_a\, d^3p_b=d^3P\,d^3Q/8$.

By introducing a  space-like unit vector $\k^\mu$ which is orthogonal to $P^\mu$ -- i.e., $\k^\mu P_\mu=0$ --  the relative momentum four-vector can be written as
 \be
 Q^\mu=\frac{(m_a^2-m_b^2)c^2}{ P^2} P^\mu+\frac{\k^\mu}{ P}\sqrt{P^4-2P^2(m_a^2+m_b^2)c^2+(m_a^2-m_b^2)^2c^4}.
\ee{A4}

As in the work \cite{KM} we shall restrict ourselves to the case where the rest masses of the particles of the constituents are not too disparate so that $m_b\approx m_a,\, (\forall\, b=1,\dots, r)$. In this case   the relative momentum four-vector and its modulus can be approximated by
 \be
 Q^\mu=Q\,{\k^\mu}, \qquad Q^2=P^2-4m_am_b,
\ee{A5}
Furthermore, in this approximation $P^\mu Q_\mu=0,$ and the  the invariant flux  reduces to
\be
F_{ba}=\sqrt{(p^\mu_ap_{b\mu})^2-m_a^2m_b^2c^4}=\frac{PQ}{2}.
\ee{A6}

In terms of the total and relative momentum four-vectors the linearized product of the distribution functions can be written as
\ben\lb{A7}
f_af_b=\frac{n_a n_b e^{-\frac{P^\mu U_\mu}{kT}}}{16\pi^2k^2 T^2c^2 m_a^2 m_b^2 K_2(\z_a)K_2(\z_b)}\left[1-\frac{\J_a^\mu}{2\p_a}\left(P_\mu+Q_\mu\right)-\frac{\J_b^\mu}{2\p_b}\left(P_\mu-Q_\mu\right)\right],
\een
thanks to (\ref{13}) and (\ref{A3a}).

 For the integration of (\ref{A1}) with respect to the relative momentum
 four-vector, the center-of-mass system is
 chosen where the spatial components of the total momentum
 four-vector vanish. By introducing the representations $(P^\mu)=(P^0,{\bf 0})$
and $(Q^\a)=(0,{\bf Q})$  we can  obtain:
 \be
 F_{ba}\frac{d^3p_a}{ p_{a0}}\frac{d^3p_b}{ p_{b0}}=\frac{1}{4\sqrt{g_0}}\frac{d^3P\,d^3Q}{P_0}.
 \ee{A8}

 Now we write the element of solid angle  as  $d\Omega=\sin\Theta d\Theta d\Phi$, where $\Theta$ and
$\Phi$ are polar angles of the spatial components of $Q^{\prime\alpha}$ with respect to
$Q^{\alpha}$ and such that $\Theta$ represents the scattering angle.
The differential cross section $\sigma_{ab}$ depend on the modulus of the relative momentum four-vector $Q$ and on the scattering angle $\Theta$, i.e, $\sigma_{ab}=\sigma_{ab}(Q,\Theta)$.
 Here we are interested in analyzing a mixture where the differential cross section is constant, which in the non-relativistic case corresponds to a mixture of hard spheres particles.
 Hence without loss of generality, we may assume that the spatial component of $Q^{\alpha}$ is in the direction of the axis
$x^3$, so that we can write $Q^{\alpha}$ and $Q'^{\alpha}$ as:
\be
(Q^{\mu})=Q\pmatrix{0\cr 0\cr 0\cr 1},
\qquad\qquad
(Q^{\prime\mu})=Q\pmatrix{0\cr \sin\Theta\cos\Phi\cr
\sin\Theta\sin\Phi\cr \cos\Theta}.
\ee{A9}

From the above considerations we can obtain the following result
\ben\lb{A10}
\int(p_a^{\prime\mu}-p_a^\mu)d\Omega=\frac{1}{2}\int(Q^{\prime\mu}-Q^\mu)d\Omega=-2\pi\,Q^\mu.
\een
We can also perform the integrations in the spherical angles of
$Q^\mu$, denoted by $\theta$ and $\phi$, i.e.
\be (Q^{\mu})=Q\pmatrix{0\cr
\sin\theta\cos\phi\cr  \sin\theta\sin\phi\cr \cos\theta},
\qquad
d^3Q=Q^2\sin\theta\,d\theta\,d\phi\,dQ=Q^2d\Omega^\star dQ,
\ee{A11}
where $d\Omega^\star= \sin\theta d\theta d\phi$ represents an element of
solid angle. Hence, we can obtain the following expressions for the integrals
\be
\int d\Omega^\star=\int_0^{2\pi}\int_0^{\pi}
\sin\theta\,d\theta\,d\phi=4\pi,\qquad \int Q^{\mu} d\Omega^\star=0,
\qquad
\int Q^{\mu}Q^{\nu}d\Omega^\star={4\pi\over 3}Q^2
\left({P^{\mu}P^{\nu}\over P^2}-g^{\mu\nu}\right).
\ee{A12}

 By using the above results the integral (\ref{A1}) can be written as
 \be
 ^3\mathcal{I}_{ab}^\mu=\frac{n_a n_b \,\sigma_{ab}}{48k^2 T^2c\, m_a^2 m_b^2 K_2(\z_a)K_2(\z_b)}\left(\frac{\J_{a\nu}}{\p_a}-\frac{\J_{b\nu}}{\p_b}\right)\int e^{-\frac{P^\mu U_\mu}{kT}}\left({P^{\mu}P^{\nu}\over P^2}-g^{\mu\nu}\right)Q^5dQ\sqrt{g_1^3g_0}\frac{d^3P}{P_0}.
 \ee{A13}

For the integration of (\ref{A13}) we consider a Lorentz rest frame where $(U^\mu)=(c/\sqrt{g_0},{\bf 0})$, a spherical
 coordinate system  and write  $d^3P=\vert{\bf P}\vert^2 d\vert{\bf P}\vert \sin\vartheta\, d\varphi\, d\vartheta$. Furthermore, we
 introduce a new variable of integration
\ben\lb{A14}
x=\sqrt{\z_a\z_b\left(\frac{Q^2}{m_am_bc^2}+4\right)},\qquad\hbox{so that}\qquad Q\,dQ=\left(\frac{kT}{c}\right)^2x \,dx
\een
where the range of this new variable is given by $2\sqrt{\z_a\z_b}\leq x<\infty$ and  write the time and spatial coordinates of
the total momentum four-vector as
\be
P_0={kT\over c} \sqrt{g_0}\,xy, \qquad\vert{\bf P}\vert={kT\over c}\frac{x\sqrt{y^2-1}}{g_1}.
\ee{A15}
In this case the element of integration becomes
\be
\sqrt{g_1^3g_0}\frac{d^3P}{P_0}=\left({kT\over c}\right)^2x^2\,y\,\sqrt{y^2-1}\,\sin\vartheta\, d\varphi\, d\vartheta\,dy,
\ee{A16}
and the range of integration of the variable $y$ reads $1\leq y<\infty$.

From the integration  over the solid angle and over the variable $y$ we can obtain the following results
\ben\lb{A17a}
&&\sqrt{g_1^3g_0}\int e^{-\frac{P^\mu U_\mu}{kT}}\frac{d^3P}{P_0}=4\pi\left({kT\over c}\right)^2xK_1(x),\\\lb{A17b}
 &&\sqrt{g_1^3g_0}\int e^{-\frac{P^\mu U_\mu}{kT}}\frac{P^\mu P^\nu}{P^2}\frac{d^3P}{P_0}=4\pi\left({kT\over c}\right)^2\left[xK_3(x)\frac{U^\mu U^\nu}{ c^2}-K_2(x)\,g^{\mu\nu}\right],
\een
so that (\ref{A13}) reduces to
\be
 ^3\mathcal{I}_{ab}^\mu=-\frac{4n_a n_b \sigma_{ab}}{3c   K_2(\z_a)K_2(\z_b)}\left(\frac{\J_{a}^\mu}{\p_a}-\frac{\J_{b}^\mu}{\p_b}\right)\int_{2\sqrt{\z_a\z_b}}^\infty \left[x^2K_3(x)-3xK_2(x)\right]
 \left[\frac{x^4}{16\z_a^2\z_b^2}-\frac{x^2}{2\z_a\z_b}+1\right]dx.
 \ee{A18}

 To perform the last integration in the variable $x$ in (\ref{A18}) we need the following integrals of Bessel functions
\ben
\int_\chi^\infty x^5 K_2(x)\,dx=\chi^4\left[\chi K_5(\chi)-6K_4(\chi)\right],\qquad
\int_\chi^\infty x^3 K_2(x)\,dx=\chi^3K_3(\chi),\\
\int_\chi^\infty x K_2(x)\,dx=\chi K_1(\chi)+2K_0(\chi),\qquad
\int_\chi^\infty x^6 K_3(x)\,dx=\chi^5\left[\chi K_6(\chi)-8K_5(\chi)\right],\\
\int_\chi^\infty x^4 K_3(x)\,dx=\chi^4 K_4(\chi),\qquad
\int_\chi^\infty x^2 K_3(x)\,dx=\chi^2 K_2(\chi)+4\chi K_1(\chi)+8K_0(\chi).
\een
Hence (\ref{A18}) becomes
\be
 ^3\mathcal{I}_{ab}^\mu=-\Upsilon_{ab}\left(\frac{\J_{a}^\mu}{\p_a}-\frac{\J_{b}^\mu}{\p_b}\right),
 \ee{33b}
 where we have introduced the following abbreviation
 \be
 \Upsilon_{ab}=\frac{16n_a n_b \sigma_{ab}}{3c   K_2(\z_a)K_2(\z_b)}\left[\frac{7}{\sqrt{\z_a\z_b}}K_3\left(2\sqrt{\z_a\z_b}\right)+\left(2+\frac{1}{\z_a\z_b}\right)K_2\left(2\sqrt{\z_a\z_b}\right)\right].
 \ee{33c}

\section{Appendix: Isotropic Schwarzschild metric}
 For the isotropic Schwarzschild metric (see e.g. \cite{Bu})
 \ben\lb{a2}
 ds^2=g_0(r)\left(dx^0\right)^2-g_1(r)\delta_{ij}dx^idx^j,\qquad
 g_0(r)=\frac{\left(1-\frac{GM}{2c^2r}\right)^2}{\left(1+\frac{GM}{2c^2r}\right)^2},\qquad
g_1(r)=\left(1+\frac{GM}{2c^2r}\right)^4,
\een
the Cristoffel symbols read
 \ben\lb{a4a}
&& \Gamma_{00}^0=0,\qquad \Gamma_{ij}^0=0,\qquad \Gamma_{ij}^k=0 \quad (i\neq j\neq k),\qquad
\Gamma_{0j}^i=0,
 \\\lb{a4b}
  &&\Gamma_{\underline{i}\,j}^{\underline{i}}=\frac{1}{2g_1(r)}\frac{d g_1(r)}{dr}\delta_{jk} \frac{x^k}{r},
 \qquad\Gamma_{0i}^0=\frac{1}{2g_0(r)}\frac{d g_0(r)}{dr}\delta_{ij}
 \frac{x^j}{r},
 \\\lb{a4d}
 &&\Gamma_{00}^i=\frac{1}{2g_1(r)}\frac{d g_0(r)}{dr}\frac{x^i}{r},\qquad
 \Gamma_{\underline{i}\,\underline{i}}^j=-\frac{1}{2g_1(r)}\frac{d g_1(r)}{dr}\frac{x^j}{r}\quad (i\neq j),
 \een
where the underlined indices are not summed and
\ben\lb{a5a}
\frac{d g_0(r)}{dr}=\frac{2GM}{c^2r^2}\frac{\left(1-\frac{GM}{2c^2r}\right)}{\left(1+\frac{GM}{2c^2r}\right)^3},\qquad
\frac{d g_1(r)}{dr}=-\frac{2GM}{c^2r^2}\left(1+\frac{GM}{2c^2r}\right)^3.
\een

\section*{Acknowledgments}
 This paper was partially supported by Conselho Nacional de Desenvolvimento Cient\'{\i}fico e Tecnol\'ogico (CNPq),  Brazil.



\begin{thebibliography}{99}



\bibitem{B}  Bernstein J, 1988 \emph{Kinetic Theory in the Expanding Universe},  (Cambridge:  Cambridge University Press)

\bibitem{AL}  Sandoval-Villalbazo A,  Garcia-Perciante A L and  Brun-Battistini D, \emph{Tolman's law in linear irreversible thermodynamics: a kinetic theory approach}, 2012 Phys. Rev. D {\bf86} 084015

 \bibitem{K} Kremer G M, \emph{Relativistic gas in a Schwarzschild metric}, 2013 J. Stat. Mech. P0000

 \bibitem{Eck}  Eckart C, \emph{The thermodynamics of irreversible processes, III.
Relativistic theory of a simple fluid}, 1940 Phys. Rev. {\bf58}, 919

\bibitem{To1} Tolman  R C, \emph{On the weight of heat and thermal equilibrium
in general relativity}, 1930 Phys. Rev. {\bf35} 904

\bibitem{To2} Tolman  R C, \emph{Temperature equilibrium in a static
gravitational field}, 1930 Phys. Rev. {\bf36} 1791

\bibitem{CK}   Cercignani C and  Kremer G M, 2002 {\it
The Relativistic Boltzmann Equation: Theory and Applications}
(Basel: Birkh\"auser)

\bibitem{BGK}  Bhatnagar P L, Gross E P and   Krook M, \emph{A model for collision
processes in gases. Small amplitude processes in charged and neutral
one-component systems}, 1954 { Phys. Rev.} {\bf 94} 511

\bibitem{WE} Welander P, \emph{On the temperature jump in a rarefied gas}, 1954
{Arkiv  f\"or Fysik} {\bf 7}, 507

\bibitem{Mar1}  Marle C, \emph{Sur l'\'etablissement des \'equations de l'
hydrodynamique des fluides relativistes dissipatifs, I. L'\'equation de
Boltzmann relativiste}, 1969  { Ann. Inst. Henri Poincar\'e}
{\bf 10} 67

\bibitem{Mar2}  Marle C, \emph{Sur l'\'etablissement des \'equations de l'
hydrodynamique des fluides relativistes dissipatifs, II. M\'ethodes de
r\'esolution approch\'ee de
l'\'equation de Boltzmann relativiste}, 1969 { Ann. Inst. Henri Poincar\'e}
{\bf 10} 127

\bibitem{AW}  Anderson J L and  Witting  H R, \emph{A relativistic relaxation-time
model for the Boltzmann equation}, 1974 { Physica} {\bf 74}, 466


\bibitem{Mc}  McCormack F J,
     \emph{Construction of linearized kinetic models for
                    gaseous mixtures and molecular gases}, 1973
     {Phys. Fluids} {\bf 16}, 2095

\bibitem{GSB}
    Garz\`o V, Santos A and  Brey J J,
     \emph{A kinetic model for a multicomponent gas}, 1989
     { Phys. Fluids} {\bf 1}, 380



\bibitem{KPS}
       Kremer G M,  Pandolfi Bianchi M and Soares  A J,
      \emph{A relaxation kinetic model for transport phenomena in a reactive flow}, 2006
      {Phys. Fluids} {\bf 18} 037104

\bibitem{KM}  Kremer G M and  Marques Jr. W, \emph{Grad's moment method for relativistic gas mixtures of Maxwellian particles}, 2013 Phys.Fluids {\bf25} 017102



\bibitem{CC} Chapman S and  Cowling T G, 1970 \emph{
The Mathematical Theory of Non-Uniform Gases, 3rd. edition}
(Cambridge: Cambridge University Press)

\bibitem{KB}   Kremer G M, 2010 \emph{
An Introduction to the Boltzmann Equation and Transport Processes  in Gases}
(Berlin: Springer)


\bibitem{St}  Stewart J M, 1971 \emph{ Non-equilibrium Relativistic Kinetic
Theory}, Lecture Notes in Physics, vol. 10 (Heidelberg: Springer)


\bibitem{Bu} Adler R,  Bazin M and  Schiffer M, 1965 \emph{Introduction to General Relativity} (New York: McGraw Hill)







\end{thebibliography}
 \end{document}